*The Core of Directed Network Problems with Quotas*

Somdeb Lahiri
School of Economics and Business Sciences,
University of the Witwatersrand at Johannesburg,
Johannesburg,
South Africa.
Email: lahiris@sebs.wits.ac.za

## *Abstract*

This paper proves the existence of non-empty cores for directed network problems with quotas and for those combinatorial allocation problems which permit only exclusive allocations.

## Introduction

Networks among a group of agents, arise very often in society as well as in economic analysis. In a network, pairs of agents are linked to each other in a symmetric relationship. Slikker and van den Nouweland [2001], Dutta and Jackson [2003] and Jackson [2004], study the problem of network formation. In a recent work, Jackson and van den Nouweland [2004] study the existence of networks that are stable against changes in links by any coalition of individuals.
However, not all interactions among individuals are of necessity symmetrical. Thus, for instance, when an agent decides to buy an object from another individual, it is not necessary that the resulting transaction, materializes in a direct exchange of objects. This problem was analyzed rigorously by Shapley and Scarf [1974], where each of a set of individuals was initially endowed with exactly one object and an allocation of the objects which could not be improved upon by any coalition of individuals by redistributing their initial endowments, was sought. An allocation such as this was called core stable and Shapley and Scarf [1974], used Gale's Top Trading Cycle Algorithm to show that a core stable allocation for such a situation would always exist.
In the situation that Shapley and Scarf[1974] analyzed, an allocation did not necessarily correspond to a network. If one agent received the object owned by a second, it did not follow that the second received the object owned by the first in return. An allocation would however correspond to what is known as a directed network. If a link was established from one agent to a second, all that it would imply is that the first agent received the object owned by the second.
The situation studied by Shapley and Scarf[1974] was characterized by two features: (a) each agent consumed exactly one object: (b) excludability in consumption. While, the first feature was perhaps only a simplifying assumption, meant largely to facilitate exposition as subsequent research as well as a later section of our paper reveals, the same cannot be said of the second feature. Excludability in consumption implies that at most one agent can consume a particular item, a characteristic associated with "private goods". There are many situations and objects which do not qualify this property. For instance, an internet server can be linked simultaneously to several other internet servers, not just one. A particular object or facility can be simultaneously used by several users, whose number does not exceed a pre-assigned quota. Such facilities or goods are akin to public goods. It is precisely such goods that we have in mind in the present context.

As in Shapley and Scarf[1974], consider a finite population of agents, each of whom is initially endowed with a single item. Each item has a capacity denoting the number of agents it can simultaneously cater too. The quota of each agent which is assumed equal to the capacity of the item she is initially endowed with, imposes an upper bound on the number of items she can consume. The requirement that the quota of an agent is equal to the capacity of her initial endowment, implies that in any directed network that would be of interest in the present context, the number of links that terminate at any agent is equal to the number of links that emanate from her. In particular a link can be a loop i.e. begin and terminate at the same agent. The problem we are concerned with here, is with the existence of a directed network which satisfies individual quotas and is core stable in the following sense: there does not exist any coalition of agents who can link up among themselves and do better than at the existing network. We show that a slight modification of the Gale's Top Trading Cycles Algorithm that was used by Shapley and Scarf[1974], proves the existence of a core stable network for every directed network problem with quotas.

A directed network problem is an example of a combinatorial allocation problem (CAP). A CAP is a resource allocation problem, in which a non-empty set of items are to be allocated across a set of agents. Agents are assumed to value bundles of items. The CAP is relevant to many interesting and important real-world applications, including scheduling, logistics and network computation domains. When the set of items is identical to the set of agents, we have a directed network problem.

An allocation of items in a CAP, where each individual is initially endowed with a distinct non-empty set of items is said to be exclusive, if no two items share an item at the allocation.

Assuming that the maximum number of items that each agent can consume at any allocation is equal to the number of items she was initially endowed with, and restricting attention to exclusive allocations, we show that a core stable allocation always exists for such CAP's. The proof of this result is very similar to the proof of the non-emptiness of the core of a directed network problem with quotas.

### **The Directed Network Problem**

Given a non-empty finite set *I* of agents, a preference relation for agent $i \in I$ is summarized by a linear order R(i) over I.

A directed network is a function A: $I \rightarrow 2^I$, where I is a non-empty finite set of agents. A directed network is said to be a network if for all $A \in \Lambda$ and $i,j \in I$: [$j \in A(i)$ implies $i \in A(j)$].

We assume that each agent has a quota which is a natural number less than or equal to the cardinality of I. Hence, a quota function is a function $q: I \to \{0,\ldots,|I|\}$.
A directed network problem with quotas is the ordered pair $E = (\Lambda(q), <R(i) / i \in I>)$, where $\Lambda(q) = \{A / A \text{ is a directed network satisfying } [\text{for all } i \in I: |A(i)| \leq q(i)]\}$.
A directed network A for $E = (\Lambda(q), <R(i)/i \in I>)$ is said to be a feasible network (or simply "feasible") if $A \in \Lambda(q)$.

The reason why there may be no directed network at which all agents exhaust their quota is illustrated by the following examples.

**Example 1**: Let $I = \{1,2,3\}$ with $q(1) = 1$, $q(2) = 3$ and $q(3) = 3$. Let $A \in \Lambda(q)$ be such that $|A(3)| = 3$. Then, $A(1) = \{3\}$, $A(3) = \{1,2,3\}$ and $1 \notin A(2)$. Thus, $|A(2)| < 3$. Thus, it is not possible for both 2 and 3 to satisfy their quotas.

**Example 2**: Let $I = \{1,2,3,4\}$ with $q(1) = 1$, $q(2) = q(3) = q(4) = 4$. Let $A \in \Lambda(q)$ be such that for at least $i \in \{2,3,4\}$, $|A(i)| = 4$. Without loss of generality suppose $|A(4)| = 4$. Thus, $A(4) = I$ and $A(1) = \{4\}$. Since $q(1) = 1$, $1 \notin A(2) \cup A(3)$. Thus, $|A(2)| < 4$ and $|A(3)| < 4$. Thus, at least two agents in $\{2,3,4\}$ must have unexhausted quotas at any $A \in \Lambda(q)$.

A feasible network A is said to be blocked by a coalition (: a non-empty subset of agents) M, if there exists a permutation $p: M \to M$ and a function $y: M \to \bigcup_{i \in M} A(i)$
such that for all $i \in M$: (i)$p(i) R(i) y(i)$; (ii) $p(i) \in I \backslash A(i)$.
An alternative way of defining the concept of blocking by a coalition would be by using the concept of a unilateral hyper-relation due to Aizerman and Aleskerov [1995]. A unilateral hyper-relation on I is a subset of $2^I \times I$.
For $i \in I$ and $(S, j) \in 2^I \times I$, we write $S \geq_i j$ if and only if either $j \in S$ or $kR(i)j$ for all $k \in S$. Clearly $\geq_i$ is a unilateral hyper-relation for all $i \in I$.
A feasible network A is said to be blocked by a coalition (: a non-empty subset of agents) M, if there exists a permutation $p: M \to M$ such that for no $i \in M$ is it the case that $A(i) \geq_i p(i)$.
A feasible network A is said to belong to the core of the directed network problem with quotas E, if it is not blocked by any coalition.
The core of E, denoted Core(E) is the set of feasible networks belonging to the core of E.

Given a list of distinct agents $i_1,\ldots,i_k$ we say that a transaction is completed along the cycle $(i_1,\ldots,i_k)$ if each $i_j \in \{i_2,\ldots,i_k\}$ receives $i_{j-1}$ and $i_1$ receives $i_k$. Thus, if $k = 1$, then after completion of transaction along the cycle, agent $i_1$ receives $i_1$.

The proof of the following theorem, which is a generalization of the one in Shapley and Scarf[1974], relies on a minor variation of the Gale's Top Trading Cycles Algorithm. Our proof itself is a modification of the one in Shapley and Scarf[1974].

Theorem 1: If E is a directed network problem with quotas, then Core(E) is non-empty.

Proof: Stage 1: Each agent i points to the agent who owns her most preferred object according to the linear order $R_i$. Since, the number of agents is finite, there exists at least one subset of agents who form a cycle, i.e. there exists a set $i_1,\ldots,i_k$ of agents, such that $i_j$ is the most preferred item of agent $i_{j-1}$ for $i_j \in \{i_2,\ldots,i_k\}$ and $i_1$ is the most preferred item of agent $i_k$. Since each agent points to exactly one agent, no two distinct cycles can share an agent. Otherwise, there would exist an agent who points to two different agents, contrary to hypothesis. Complete the transaction along each such cycle.
Each agent who does not get an object she had pointed to, was not part of a cycle. Each agent who received an object at this stage, strikes that object off from her linear order.
Each agent who received an object up until this stage, reduces her quota by one, to obtain revised quotas. Any agent whose quota has been reduced to zero, withdraws from the procedure. If in the process all agents withdraw from the procedure, the procedure terminates. Otherwise the procedure moves to Stage 2, with participating agents being only those agents who either did not receive an item at Stage 1 or whose revised quota after stage 1 is positive. No agent whose quota is incomplete is removed from the linear order of any participating agent.
Each agent who participate in the subsequent stage removes from her linear order all agents who have exhausted their quota. Each agent who received an object at Stage 1 and proceeds to participate in the subsequent stage, removes from her linear order the (owner of) the item she received at Stage 1.
Stage 2: Repeat Stage 1, among the participating agents. (This may involve an agent pointing to an agent she had at stage 1). Each agent who received an object up until this stage, reduces her quota by one, to obtain revised quotas.
Repeating the process at most a finite number of times, we arrive at a stage where either all agents have filled their quota, or the agents who have not filled there quota, have by now struck all agents off their list.
The procedure terminates now with the directed network A being defined such that for all $i \in I$, A(i) is the set of items currently in the possession of agent i.
We claim that A belongs to the core of E. A is clearly feasible.
Suppose there is a coalition M which blocks A. Thus there exists a permutation p: $M \to M$ and a function y: $M \to \bigcup_{i \in M} A(i)$ such that for all $i \in M$: (i) p(i) R(i) y(i); (ii) p(i) $\in$ I\A(i).
Without loss of generality let an agent in M whose quota was exhausted first among all agents in M, be denoted 1. If agent 1's quota was exhausted at the first stage of the procedure, she clearly got her best item and therefore could not be part of a blocking coalition. Hence no agent whose quota was exhausted at the first stage would be part of a blocking coalition. If agent 1's quota was exhausted at stage 2, then the only agents that she could form a blocking coalition with, must have exhausted their quota in stage 1. Since agents who exhausted their quota in stage 1 cannot belong to M, it is not possible for agent 1 to belong to M either. Thus, no agent whose quota was exhausted at stage 2 can belong to M. Proceeding thus, we see that if agent 1's quota was exhausted at stage k, then the owners of the items she could form a blocking coalition with must have exhausted their quota at a previous stage. Since agent 1 is assumed to be among the first to exhaust her quota among the agents in M, M cannot be a blocking coalition. This contradiction establishes the non-emptiness of the Core(E). Q.E.D.

The purpose of requiring the termination rule in the above procedure to permit agents whose quota may have remained unexhausted may once again be illustrated by the following example.

**Example 3**: Let I = {1,2} and suppose q(1) = 1 where as q(2) = 2. Let E be a directed network problem where agent 1 prefers 1 to 2. If A belongs to Core(E), then A(1) = {1}. Thus, whatever be the preference of agent 2, A(2) = {2}. In fact this would be the unique feasible network in Core(E). Clearly, the quota of agent 2 remains unexhausted at A.
However, thee feasible network where agent 1 gets 2 and agent 2 gets both 1 and 2, exhausts the quota of all agents. This network is blocked by agent 1 and hence does not belong to the core.

**Note**: Suppose the directed network A obtained in the proof of theorem 1 above, was the outcome of a procedure that terminated at stage K ≥ 1. For all agents i∈I, who received item j at stage K, let $p_K = p_i(j) = 1$.

If K > 1, then having defined $p_k$ for stages K, K-1,…, L < 1, define $p_{L-1} = \sum_{k=L}^{K} p_k + 1$.

For all i ∈I, who receive an item j at stage k∈ {L-1,…,K}, let $p_i(j) = p_k$.
For i∈I and j∈A(i), $p_i(j)$ may be interpreted as a personalized price of item j to agent i.
For all j∈I, let p(j) = min {$p_i(j)$ / j∈A(i), i∈I}. For i∈I and j∈I\A(i), let $A^{-j}(i) = \{h∈A(i)/ jR(i)h\}$.
The pair (A, <$p_i(j)$/ j∈A(i), i∈I> satisfies the following property: (i) for all i∈I and j∈I\A(i) with $A^{-j}(i) ≠ \phi$: $p(j) > \sum_{h \in A^{-j}(i)} p_i(h)$; (ii) for all i ∈I, j ∈A(i) and h ∈I\A(i), [ hR(i) j] implies [ p(h) > $p_i(j)$]; (iii) for all i∈I: $\sum_{\{j / j \in A(i)\}} p_i(j) = \sum_{\{j / i \in A(j)\}} p_j(i)$.

(i) says that for any agent i and any item j not belonging to A(i), the total payment that agent i makes for items she does not prefer to item j, is less than the least personalized price paid for item j. Now, (ii) follows from (i) since all personalized prices computed above, are strictly positive. (iii) says that, given any agent i, the sum of payments made by i is equal to the sum of payments received by i .

## A Combinatorial Allocation Problem and Its Non-empty Core

In a combinatorial allocation problem (CAP) there is a non-empty finite set $G$ of discrete items and a non-empty finite set $I$ of agents. Each agent $i \in I$, is initially endowed with a non-empty set of items S(i), such that: (a) $\bigcup_{i \in I} S(i) = G$; (b) for all i,j∈I, with i≠j: S(i)∩S(j) = $\phi$.

A preference relation for agent i∈I is summarized by a linear order R(i) over G.
An allocation is a function A: $I \rightarrow 2^G$.
For all i∈I, A(i) is the bundle received by agent i.
An allocation A is said to be exclusive if for all i,j∈I with i ≠ j: A(i) ∩ A(j) = $\phi$.
Let q: I→ N(: the set of natural numbers) be such that for all i∈I, q(i) = |S(i)|, i.e. the cardinality of S(i).
Let $\Lambda$(q) = {A/ A is an exclusive allocation satisfying |A(i)| = q(i) for all i∈I}.
An allocation A is said to be feasible if A∈$\Lambda$(q).
For i∈I and (S, a) $\in 2^G \times G$, we write $S \geq_i a$ if and only if either a∈S or bR(i)a for all b∈S.
Clearly $\geq_i$ is a unilateral hyper-relation for all i∈I.
A feasible allocation A is said to be blocked by a coalition (: a non-empty subset of agents) M, if there exists a permutation p: M→ M and a function x: M→$\bigcup_{i \in I} S(i)$ such that for no i∈M is it the case that A(i) $\geq_i$ x(p(i)).
A feasible allocation A is said to belong to the core of the CAP, if it is not blocked by any coalition.
The core of the above CAP, denoted C* is the set of feasible networks belonging to its core.

Given a list of distinct agents $i_1,\ldots,i_k$ and a set of items $x(i_1),\ldots, x(i_k)$ with x(i)∈S(i) for all i∈ $\{i_1,\ldots,i_k\}$ we say that a transaction is completed along the cycle $(i_1,\ldots,i_k)$ if each $i_j \in \{i_2,\ldots,i_k\}$ receives $x(i_{j-1})$ and $i_1$ receives $x(i_k)$. Thus, if k = 1, then after completion of transaction along the cycle, agent $i_1$ receives $x(i_1)$.

The proof of the following theorem, which is again a generalization of the one in Shapley and Scarf[1974], is almost identical to the proof of our Theorem 1.

Theorem 2: Given a CAP as defined above, C* is non-empty.

Proof: The only respects in which the proof here differs from the proof of Theorem 1, is that every time a transaction is completed along a cycle, all the items involved in the transaction are struck off from the list of all agents who participate in the subsequent stage of the procedure and the procedure stops when the quotas of all agents have been exhausted. Let an allocation A be the outcome of the procedure thus defined.
We claim that A belongs to the C*. A is clearly feasible.

Suppose there is a coalition M which blocks A. Thus there exists a permutation p: M→ M and functions x: M→$\bigcup_{i \in I} S(i)$, y: M → $\bigcup_{i \in M} A(i)$ such that for all i∈M:
(i) x(p(i)) R(i) y(i); (ii) x(p(i))∈G\A(i).

Without loss of generality let an agent in M whose quota was exhausted first among all agents in M, be denoted 1. If agent 1's quota was exhausted at the first stage of the procedure, she clearly got her best item and therefore could not be part of a blocking coalition. Hence no agent whose quota was exhausted at the first stage would be part of a blocking coalition. If agent 1's quota was exhausted at stage 2, then the only agents that she could form a blocking coalition with, must have exhausted their quota in stage 1. Since agents who exhausted their quota in stage 1 cannot belong to M, it is not possible for agent 1 to belong to M either. Thus, no agent whose quota was exhausted at stage 2 can belong to M. Proceeding thus, we see that if agent 1's quota was exhausted at stage k, then the owners of the items she could form a blocking coalition with must have exhausted their quota at a previous stage. Since agent 1 is assumed to be among the first to exhaust her quota among the agents in M, M cannot be a blocking coalition. This contradiction establishes the non-emptiness of C*.
Q.E.D.